\documentclass[conference]{IEEEtran}
\usepackage{blindtext, graphicx}
\usepackage{multirow}
\usepackage{amssymb,mathtools}
\usepackage{color}
\usepackage{algorithm}
\usepackage[algo2e]{algorithm2e}
\usepackage{subfigure}
\begin{document}

\title{User Transmit Power Minimization through Uplink Resource Allocation and User Association in HetNets}

\author{\IEEEauthorblockN{Umar Bin Farooq\IEEEauthorrefmark{1},
Umair Sajid Hashmi\IEEEauthorrefmark{2},
Junaid Qadir\IEEEauthorrefmark{1}, 
Ali Imran\IEEEauthorrefmark{2} and
Adnan Noor Mian\IEEEauthorrefmark{1}}
\IEEEauthorblockA{\IEEEauthorrefmark{1}Information Technology University, Punjab, Pakistan\\ Email: (mscs16019,junaid.qadir,adnan.noor)@itu.edu.pk}
\IEEEauthorblockA{\IEEEauthorrefmark{2}School of Electrical and Computer Engineering, University of Oklahoma, Tulsa, OK, USA\\
Email: (umair.hashmi,ali.imran)@ou.edu}}

\maketitle

\begin{abstract}
The popularity of cellular internet of things (IoT) is increasing day by day and billions of IoT devices will be connected to the internet. Many of these devices have limited battery life with constraints on transmit power. High user power consumption in cellular networks restricts the deployment of many IoT devices in 5G. To enable the inclusion of these devices, 5G should be supplemented with strategies and schemes to reduce user power consumption. Therefore, we present a novel joint uplink user association and resource allocation scheme for minimizing user transmit power while meeting the quality of service. We analyze our scheme for two-tier heterogeneous network (HetNet) and show an average transmit power of -2.8 dBm and 8.2 dBm for our algorithms compared to 20 dBm in state-of-the-art Max reference signal received power (RSRP) and channel individual offset (CIO) based association schemes. 
\end{abstract}


\IEEEpeerreviewmaketitle

\section{Introduction}
Internet of Things (IoT) is finding a wide range of applications in smart cities, sensor network, healthcare, industrial automation, and agriculture. This is causing an expeditious increase in the number of IoT devices and it is expected that 20 billion devices will be connected to the Internet by 2020 \cite{bahga2014internet}. IoT devices have different throughput, latency, and battery-related constraints. The transmit power of battery constrained devices is usually limited and the number of these battery constrained devices are bound to increase due to the popularity of IoT applications. Fifth generation (5G) network is expected to solve complex challenges of current communications systems by providing intelligent strategies for smooth integration of IoT devices. 

Telecommunication research community is currently exploring new schemes such as communication spectrum at higher frequencies, new physical layer techniques and network densification to meet the rising demand for availability and throughput. Network densification remains one of the most promising themes for 5G \cite{bhushan2014network}. Network densification creates a HetNet by introducing small base stations (BS) with traditional macro BS. Although HetNet brings users and BS closer providing higher quality links and frequency reuse, it poses new challenges for the research community. Uplink-downlink asymmetry (in terms of coverage, channel quality, transmit power, and hardware limitations) is one of the pressing challenges of HetNets \cite{liu2016user}.

The uplink-downlink asymmetry makes an optimal user association for uplink or downlink non-optimal for the other. Hence, uplink-downlink separation to rectify this asymmetry problem was proposed \cite{mohamed2016control}. Users can meet both uplink and downlink objectives by associating to different BS for uplink and downlink transmission in this separated architecture. Singh et al. \cite{singh2015joint} studied uplink-downlink separation and showed that the path loss based association is optimal for uplink rate. They also showed better uplink-downlink rate coverage in a decoupled association. This decoupled architecture provides the liberty to optimize uplink association separately from downlink association. The uplink association can be optimized to minimize uplink user transmit power. The battery life of power constrained devices will significantly improve from this reduction in transmit power. In addition, many new devices with more rigid limitations on transmit power will be able to communicate through cellular networks.

In this work, we exploit decoupled uplink-downlink architecture and present an uplink user association for user transmit power minimization. Our strategy opportunistically exploits unused spectrum at a BS and allocates more than the minimum bandwidth to reduce transmit power of the user.  We define residual bandwidth at a BS as the bandwidth of the BS not assigned to any user and is available for allocation to new users. Our association scheme considers residual bandwidth in addition to the path loss to serving BS. The BS with more residual bandwidth can provide more spectrum to the user and hence the user can reduce the transmit power for same quality of service (QoS) requirements.

\subsection{Contributions of this Paper}
The major contributions of this paper are listed as follows:

\begin{itemize}
  \item A novel uplink user association scheme in which a user is assigned to the BS which minimizes the transmit power. The transmit power is formulated as a function of signal path loss from user to BS and bandwidth allocated to the user. 
  \item We present resource allocation schemes to effectively utilize the residual bandwidth at the BS. We opportunistically allocate more resources to a user for lower uplink transmit power.
  \item A comparison of our scheme with state of the art maximum RSRP and CIO based association scheme shows significantly lower transmit power in our proposed strategy. 
\end{itemize}

The rest of paper is organized as follows: Section \ref{sec:relatedWork} provides a brief summary of related work. In section \ref{sec:systemModel}, we discuss system model and formulate the problem. Section \ref{sec:methodology} presents joint resource allocation and association methodology. Section \ref{sec:simulation} gives the simulation results and section \ref{sec:conclusion} concludes the paper.

\section{Related Work}
\label{sec:relatedWork}
Several user association schemes for 5G are proposed in the recent literature. A very comprehensive survey of user association schemes for HetNets, millimeter wave, massive MIMO and energy harvesting networks was presented in \cite{liu2016user}. Several uplink user association schemes used game theory to meet the optimization objective. Ha et al. \cite{ha2014distributed} explored the application of game theory in user association and used non-cooperative game theory for the uplink association. They presented a distributed association scheme with power control. They also designed a hybrid power control algorithm to meet the user SINR requirements for a two-tier HetNet. Saad et al. \cite{saad2014college} formulated the problem of uplink user association as a college admission game and the game was solved using matching theory and coalitional games. The ranking in the game was done using packet success rate, delay, and cell range expansion while maintaining the QoS requirements of the user. An uplink user association scheme with energy-efficient resource allocation meeting the minimum Quality of Service (QoS) requirements was presented by Pervaiz et al. \cite{pervaiz2013joint}. They explored the energy efficiency (EE) vs QoS tradeoffs for a two-tier Orthogonal Frequency Division Multiplexing (OFDM) HetNet. 

Recent work in user association has also explored joint optimization of uplink and downlink objectives. Chen et al. \cite{chen2012joint} presented a joint uplink and downlink association for HetNets. The association scheme jointly maximizes downlink system capacity and minimizes uplink transmit power of the user. Luo et al. \cite{luo2015downlink} presented a joint uplink and downlink association and beamforming for energy efficiency in C-RAN. They converted the joint association problem into an equivalent downlink problem with two sub-problems. Liu et al. \cite{liu2014joint} also presented a joint uplink-downlink association for energy efficiency. They formulated the association as a Nash Bargaining problem for both uplink-downlink energy efficiency. The user association problem for future deployment of cellular networks in mmWave scenarios is also discussed in recent literature \cite{cetinkaya2017user}, \cite{zhang2017energy}, \cite{al2017planning}.

\section{System Model and Problem Description}
\label{sec:systemModel}

We consider a two-tier uplink HetNet in which macrocell network is overlaid with small cells. There is at least one randomly deployed small BS in each macrocell. We assume each small BS is connected to high capacity wired backhaul. The same frequency band is used by both small and macro BS and the frequency reuse factor is 1. The interference for any user in a cell is considered from users in the neighboring cells as well as the users from the same cell. Each user $u$ has a capacity requirement $C_u$ and a maximum power threshold. To model the users with diverse capacity requirements, $C_u$ follows a uniform distribution between [0, $maxCapacity$]. The maximum power threshold is identical for all the users. The minimum bandwidth $\eta_u^c$ required by user $u$ from a BS $c$ for the maximum power threshold can be computed using Shannon equation as following:

\begin{equation}
    \eta_u^c = \frac{C_u}{log_2(1+\gamma_u^c) }
\label{eq:PRBperUser}
\end{equation}

where $\gamma_u^c$ is the SINR of user $u$ associated with BS $c$. The residual bandwidth $\eta_c$ at a BS can be computed by subtracting the sum of bandwidth allocated to all users associated with a BS $c$ from the total bandwidth $\epsilon_c$ at the BS. The residual bandwidth $\eta_c$ for the BS $c$ is computed as follows:

\begin{equation}
    \eta_c = \epsilon_c - \sum\limits_{U_c} \frac{C_u}{log_2(1+\gamma_u^c) }
\label{eq:PRBperUser}
\end{equation}

$U_c$ is the set of all active users associated with BS $c$. A set $U_u$ contains all the uplink users interfering with the user $u$. The uplink SINR $\gamma_u^c$ for user $u$ connected to base station $c$ is given by: 

\begin{equation}
    \gamma_u^c = \frac{ P_t^u G_u G_u^c \delta a (d_u^c)^{-\beta} }{K + \sum\limits_{\forall i\in U_u} P_t^i G_i G_u^i \delta a (d_i^c)^{-\beta} }
\label{eq:SINR}
\end{equation}

where $P_t^u$ is the transmit power of user $u$, $G_u$ and $G_i$ are the UE gains, $G_u^c$ is the gain from BS to UE, $\delta$ is signal shadowing, $a$ is path loss constant, $d_u^c$ and $d_i^c$ are the distances from user $u$ and interfering user $i$ to BS $c$ respectively, $\beta$ is the path loss exponent, $K$ is the thermal noise power. We define available bandwidth $BW_u^c$ as resources that a BS $c$ decides to allocate for a user $u$. $BW_u^c$ is different from $\eta_u^c$ and can vary from $\eta_u^c$ to $\eta_c$. The capacity $C_u$ for user $u$ from BS $c$ is:

\begin{equation}
    C_u =  BW_u^c log_2 (1 + \gamma_u^c) 
\label{eq:capacity}
\end{equation}

The transmit power of the user $u$ to communicate with BS $c$ can be found by replacing the value of SINR in equation \ref{eq:capacity}. The transmit power comes out to be:

\begin{equation}
    P_t^c = \bigg(2^{\frac{C_u}{BW_u^c}} - 1\bigg) \frac{ K + \sum\limits_{\forall i\in U_u} P_t^i G_i G_u^i \delta a (d_i^c)^{-\beta} }{ G_u G_u^c \delta a (d_u^c)^{-\beta} }
\label{eq:power}
\end{equation}

The UE transmit power in equation \ref{eq:power} can be optimally chosen by associating the user with the BS which can provide more bandwidth to the user meeting the user capacity requirements. The problem formulation is given as:

\begin{equation}
\begin{aligned}
& \underset{BW_u^c,d_u^c}{\text{minimize}}
& & \mathrm{\scriptstyle \Bigg( \sum\limits_{C} \bigg(\sum\limits_{U_c} { \Big(2^{\frac{C_u}{BW_u^c}} - 1\Big) \frac{ K + \sum\limits_{\forall i\in U_u} P_t^i G_i G_u^i \delta a (d_i^c)^{-\beta} }{ G_u G_u^c \delta a (d_u^c)^{-\beta} } } \bigg) \Bigg)} \\
& \text{subject to}
& & BW_u^c  \leq  \epsilon_c, \\
&&& \displaystyle \gamma_u^c  \geq  0dB.
\end{aligned}
\label{eq:optimization}
\end{equation}


\section{Resource Allocation and User Association Methodology}
\label{sec:methodology}
The state of the art user association in cellular networks is based on Max RSRP and CIO. In the Max. RSRP association scheme, the user is associated with the BS offering maximum received power; whereas for the CIO based schemes, an offset is added to small BS to offload the traffic from macro to small BS. Macro BS are more heavily loaded due to higher transmit power in Max RSRP based association \cite{siomina2012load}. Although CIO based association solves the problem of load imbalance to some extent, both these strategies do not take into account the residual bandwidth at the BS. A BS allocates the minimum required bandwidth for maximum power constraint of the user in these schemes. So, the user transmits at the maximum power threshold for most of the time. The user battery life can be increased further if the user can opportunistically transmit at a lower power than the maximum power threshold. In our scheme, the uplink association is a function of available bandwidth at the BS for the user and the received power. We allocate more bandwidth than the minimum to the user whenever BS has residual bandwidth greater than $\eta_u^c$. The user then allocate an $AssociationScore$ to each BS as described in equation \ref{eq:associationScore}. The association score takes into account both the distance-dependent channel to a BS and the allocated bandwidth to the user by a BS. We associate the user to the base station that minimizes the $AssociationScore$. 

\begin{equation}
    Association Score = \bigg(2^{\frac{C_u}{BW_u^c}} - 1\bigg)^\alpha * \bigg(\frac{ 1 }{ G_u G_u^c \delta a (d_u^c)^{-\beta} }\bigg)^{1-\alpha}
\label{eq:associationScore}
\end{equation}

where $\alpha$ is user association exponent. The $AssociationScore$ in  equation \ref{eq:associationScore} is dependent on three variables---the bandwidth that BS $u$ can allocate to user $c$, the power loss from BS $u$ to $c$ and the user association exponent. The association exponent can be varied to change the importance of residual bandwidth for the user association. There can be various methods to choose the available bandwidth for a user from residual bandwidth at a BS. A central optimal solution for equation \ref{eq:optimization} is computationally very expensive. So, we present semi-distributive and distributive schemes to approximate the gains of equation \ref{eq:optimization}. 

\subsection{Semi-Distributive Association}
\label{sec:allocation1}
Semi-distributive association uses both BS and users in a distributive manner. This association is done in two steps---optimal resource allocation at BS and optimal association at each user.

\begin{algorithm}[H]
\caption{Semi-distributive Resource Allocation and User Association Algorithm}

 \For{each user u}{
    $(a)$ Find the minimum required bandwidth $\eta_u^c$ to meet $C_u$ for all BSs as described in equation \ref{eq:PRBperUser};\\
    $(b)$\\
    \For{each candidate BS $c$}{
        compute $x_c$ which minimizes the average transmit power of user $c$;\\
        $BW_u^c = x_c*\eta_u^c$;
    }
    $(c)$ Compute $AssociationScore$ described in equation \ref{eq:associationScore} for each BS;\\ 
    $(d)$ Associate user $u$ to the BS $c$ with minimum $AssociationScore$;\\
    $(e)$ Allocate $BW_u^c$ of the serving BS $c$ to user $u$\\
    $(f)$ Update the Bandwidth of other users connected to BS c according to step $(b)$;\\
 }

\label{algo:UserAssociationAlgorithm1}
\end{algorithm}

The semi-distributive resource allocation and user association described in algorithm \ref{algo:UserAssociationAlgorithm1} has two major steps. In the first step, a user broadcasts its capacity requirements and maximum power threshold to all BS in the coverage range. Each BS calculates the minimum bandwidth $\eta_u^c$ required for the user which meets the capacity and power threshold constraints. All BS with $\eta_u^c$ less than the total bandwidth of the BS are selected as candidates for the association. Each candidate BS $c$ then finds optimal resources which minimize the average power of all users connected to $c$. $c$ assigns the optimal chunk of bandwidth to the new user which minimizes the average transmit power of all users connected to the BS. All BS satisfy the constraints of equation \ref{eq:optimization} and attempt to divide 100\% bandwidth among all the active users. Each candidate base station performs an exhaustive search to find $x_c$ in equation \ref{eq:BWAllocation1} and chooses the value which minimizes the average power consumption of all the users connected to the BS.

\begin{equation}
  BW_u^c = x_c*\eta_u^c
  \label{eq:BWAllocation1}
\end{equation}

In the second step, the user $u$ finds the optimal BS for the association which minimizes the transmit power. The value of $BW_u^c$ is communicated from all candidate BS to the user. The user then calculates the $AssociationScore$ for each BS using equation \ref{eq:associationScore}. Finally, the user associates to the BS with minimum $AssociationScore$.

\subsection{Distributive Association}
\label{sec:allocation2}
In distributive association, both resource allocation and association decisions are made at the user. A user calculates the minimum bandwidth $\eta_u^c$ for each BS given the capacity requirements and maximum power threshold. The user then considers all those BS with residual bandwidth greater than the minimum bandwidth as a candidate for the association. Each candidate BS sends residual bandwidth $\eta_c$ to the user. The user chooses the available bandwidth $BW_u^c$ for itself according to the rule described in equation \ref{eq:BWAllocation2}.

\begin{algorithm}[H]
\caption{Distributive Resource Allocation and User Association Algorithm }
 \For{each user u}{
    $(a)$ Find the minimum required bandwidth $\eta_u^c$ to meet $C_u$ for all BSs as described in equation \ref{eq:PRBperUser};\\
    $(b)$ Choose all BSs as candidates for which minimum bandwidth is less than the maximum bandwidth of the BS i.e. $\eta_u^c < \epsilon_c$;\\ 
    $(c)$\\
    \For{each candidate BS $c$}{
        \eIf{$ \eta_c > (2*\eta_u^c) $}{
            $BW_u^c = 2*\eta_u^c$
        }{
            $BW_u^c = \eta_u^c$\;
        }
    }
    $(d)$ Compute $AssociationScore$ described in equation \ref{eq:associationScore} for each BS;\\ 
    $(e)$ Associate user $u$ to the BS $c$ with minimum $AssociationScore$;\\
    $(f)$ Allocate $BW_u^c$ of the serving BS $c$ to user $u$;\\
 }
\label{algo:UserAssociationAlgorithm2}
\end{algorithm}

\begin{equation}
  BW_u^c =
  \begin{cases}
    (2*\eta_u^c) & \text{if $ \eta_c > (2*\eta_u^c) $} \\
    \eta_u^c & \text{if $ \eta_u^c < \eta_c < (2*\eta_u^c) $ }
  \end{cases}
  \label{eq:BWAllocation2}
\end{equation}

The available bandwidth that a BS can allocate to a user is twice the minimum bandwidth if the residual bandwidth at the BS is greater than twice the minimum bandwidth. If the residual bandwidth at BS is less than twice the minimum bandwidth, the available bandwidth is considered to be the minimum bandwidth. The user then calculates the $AssociationScore$ described in equation \ref{eq:associationScore} for each candidate BS using the $BW_u^c$ of \ref{eq:BWAllocation2}. The user connects to the BS with maximum $AssocationScore$. Algorithm \ref{algo:UserAssociationAlgorithm2} presents the distributive resource allocation and user association scheme.

\section{Simulation and Results}
\label{sec:simulation}
In this section, we present the simulation evaluation of the user association scheme. We also compare our user association scheme with max. RSRP and CIO based schemes in terms of user transmit power and other key performance indicators (KPIs).

\begin{table}[!h]
\centering
\caption{Description of Simulation Parameters}
\label{table:simulation}
\begin{tabular}{|l|l|}
\hline
\textbf{Parameter Description}                       & \textbf{Value}                 \\ \hline
Number of Macro BS                          & 7                     \\ \hline
Number of Sectors per Macro BS              & 3                     \\ \hline
Number of Users per sector                      & 25                    \\ \hline
System Bandwidth                            & 10 MHz                \\ \hline
Maximum User Transmit Power                 & 20 dBm                \\ \hline
\multirow{2}{*}{User Capacity Requirements} & Uniformly distributed \\ 
                                            & between 0 and 2 kHz   \\ \hline
Transmission Frequency                      & 2 GHz                 \\ \hline
Inter-cite Distance of Macro BS             & 500m                  \\ \hline
Macro BS Height                             & 25m                   \\ \hline
Small BS Height                             & 10m                   \\ \hline
Network Topology                            & Hexagonal             \\ \hline
Association Exponent                        & 0.5                   \\ \hline
User Noise Figure                        & 7 dB                   \\ \hline
Base Station Noise Figure                       & 5 dB                   \\ \hline
\end{tabular}
\end{table}

\subsection{Simulation Setup}
We employ an LTE 3GPP standard network topology \cite{3gpp} with macro cells overlaid with small cells. We deploy and simulate a two-tier HetNet with 7 macro BS in MATLAB R2014a. The simulation parameters are summarized in Table \ref{table:simulation}. Small BS are distributed in each sector of macrocell with uniform density. A fraction of both indoor and outdoor UEs are concentrated near small base stations to model hotspot scenarios. The ratio of indoor to outdoor users is 4:1. 

\subsection{Simulation Results}
A comparative analysis of semi-distributive and distributive association schemes with Max SINR and CIO based schemes using Monte Carlo simulations is presented. The performance sensitivity by varying the number of small BS per sector is also analyzed. We also discuss an inherent uplink transmit power vs spectrum efficiency tradeoff in our scheme.

\begin{figure}[!ht]
    \centering
    \subfigure[One small BS per sector]{\includegraphics[width=0.49\textwidth]{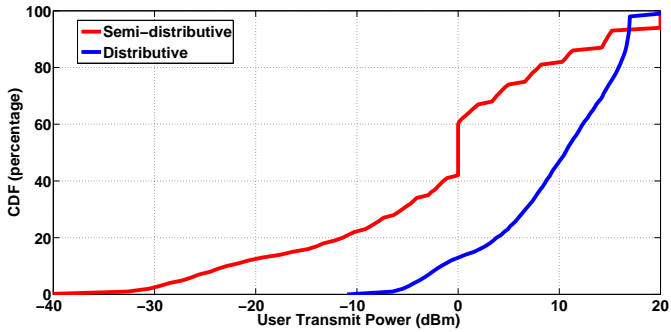}\label{fig:power1SmallBS}}
    \subfigure[Four small BS per sector]{\includegraphics[width=0.49\textwidth]{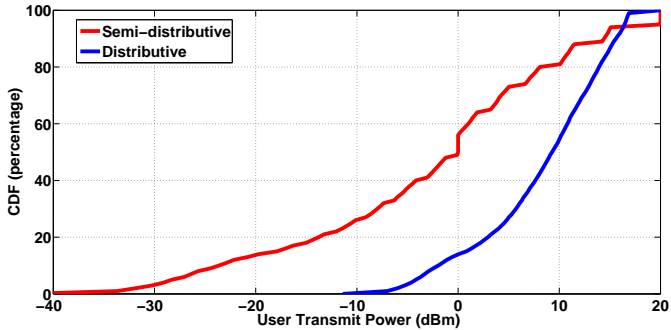}\label{fig:power4SmallBS}}
    \caption{Uplink transmit power of user in semi-distributive and distributive association schemes with. The transmit power of each user is 20 dBm in Max RSRP and CIO based association}
    \label{fig:power}
\end{figure}

\textbf{Comparison of Uplink Transmit Power:} The uplink transmit power is maximum power threshold (20 dBm) in Max RSRP and CIO based techniques where the user gets the minimum bandwidth $\eta_u^c$ from the BS. However, the user can exploit more residual bandwidth at BS in our schemes. Both semi-distributive and distributive association opportunistically allocate more bandwidth than $\eta_u^c$ to a user. Fig. \ref{fig:power1SmallBS} and \ref{fig:power4SmallBS} show a comparison of the user transmit power for semi-distributive and distributive association with 1 small BS and 4 small BS per sector. The average user transmit power in Max RSRP and CIO based association is 20 dBm for all the users while the average transmit power in the semi-distributive algorithm and distributive association algorithm decreases to -1.9 dBm and 9.4 dBm respectively for 1 small BS per sector. The average transmit power is -2.8 dBm and 8.2 dBm for semi-distributive and distributive respectively with 4 small BS per sector. The user transmit power in our algorithms decreases because we assign a weight to residual bandwidth at a BS while deciding the optimal BS for association. BS with more residual bandwidth can assign more resources to the user and hence user can transmit at a lower power to meet the same capacity requirements. 

\begin{figure}[!ht]
    \centering
    \subfigure[One small BS per sector]{\includegraphics[width=0.5\textwidth]{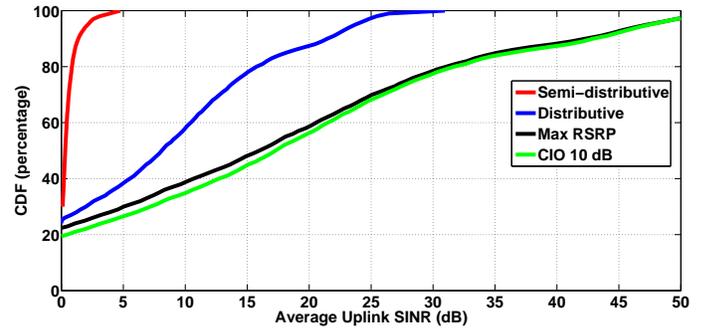}\label{fig:SINR1SmallBS}}
    \subfigure[Four small BS per sector]{\includegraphics[width=0.5\textwidth]{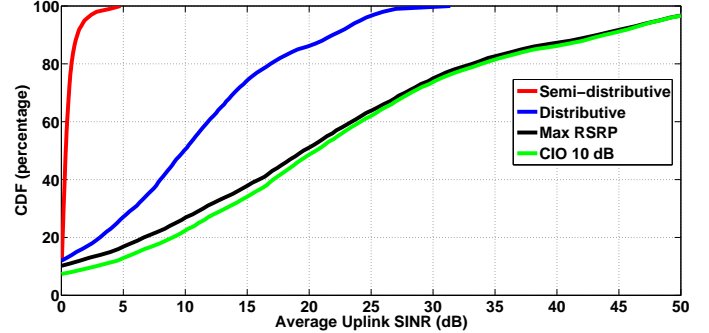}\label{fig:SINR4SmallBS}}
    \caption{Average uplink SINR of semi-distributive and distributive algorithms compared to Max RSRP and CIO based association.}
    \label{fig:SINR}
\end{figure}

\textbf{SINR comparison:} The SINR in our association algorithms is expected to decrease due to lower transmit power of the user. However, the decrease in SINR does not translate to degradation in the quality of service to the user. According to the Shannon equation, more bandwidth allocation to the user makes up for the adverse effects of poorer SINR. Hence, the capacity requirement of the user is always met as we assign more bandwidth to the user. Fig. \ref{fig:SINR1SmallBS} and \ref{fig:SINR4SmallBS} show CDFs of average uplink SINR for semi-distributive, distributive, Max RSRP and CIO based algorithm with 1 small BS and 4 small BS per sector. The uplink SINR is poor for semi-distributive and distributive algorithms compared to Max RSRP and CIO based association as expected.

\begin{figure}[!ht]
    \centering
    \subfigure[One small BS per sector]{\includegraphics[width=0.5\textwidth]{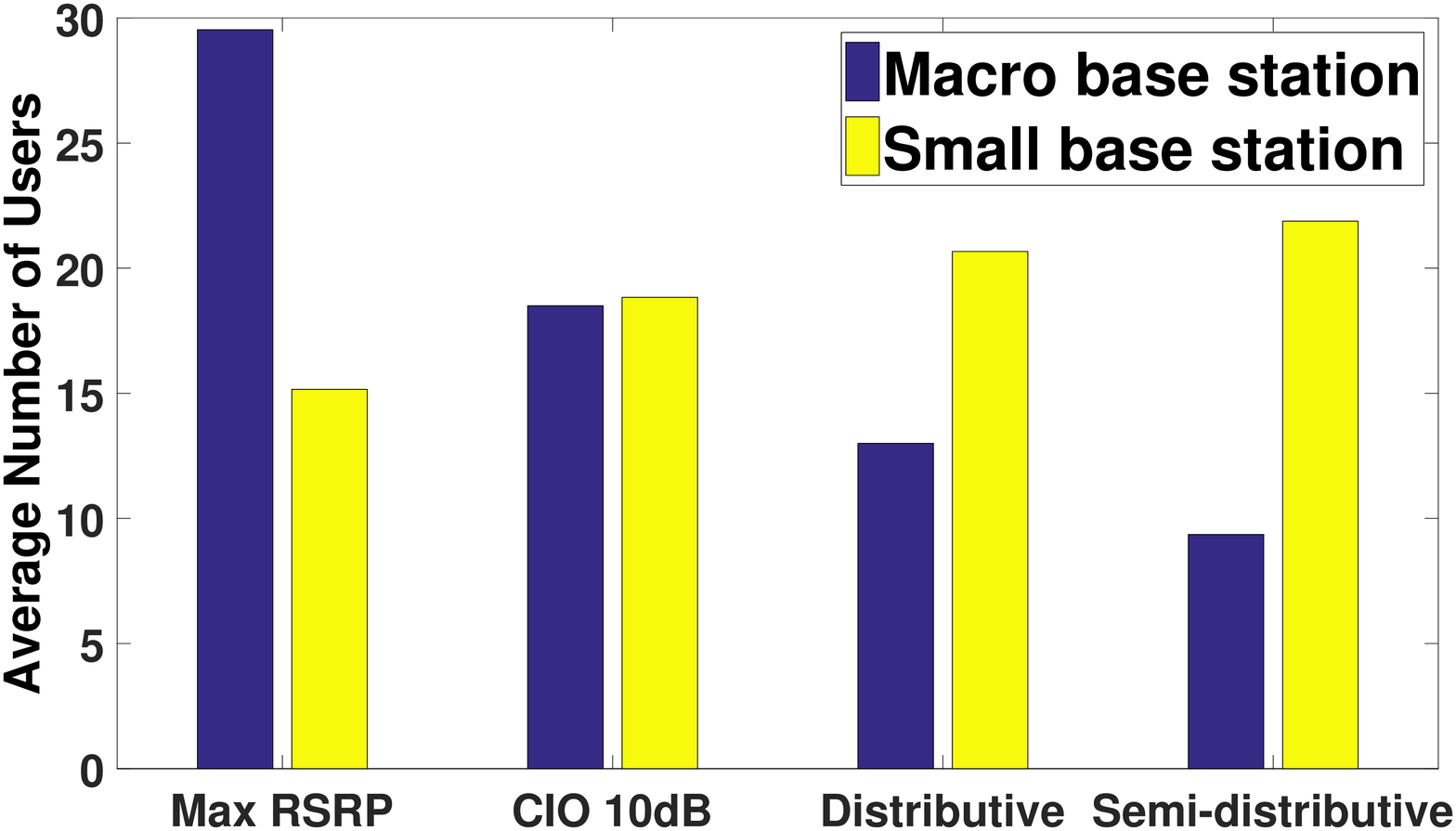}\label{fig:load1SmallBS}}
    \subfigure[Four small BS per sector]{\includegraphics[width=0.5\textwidth]{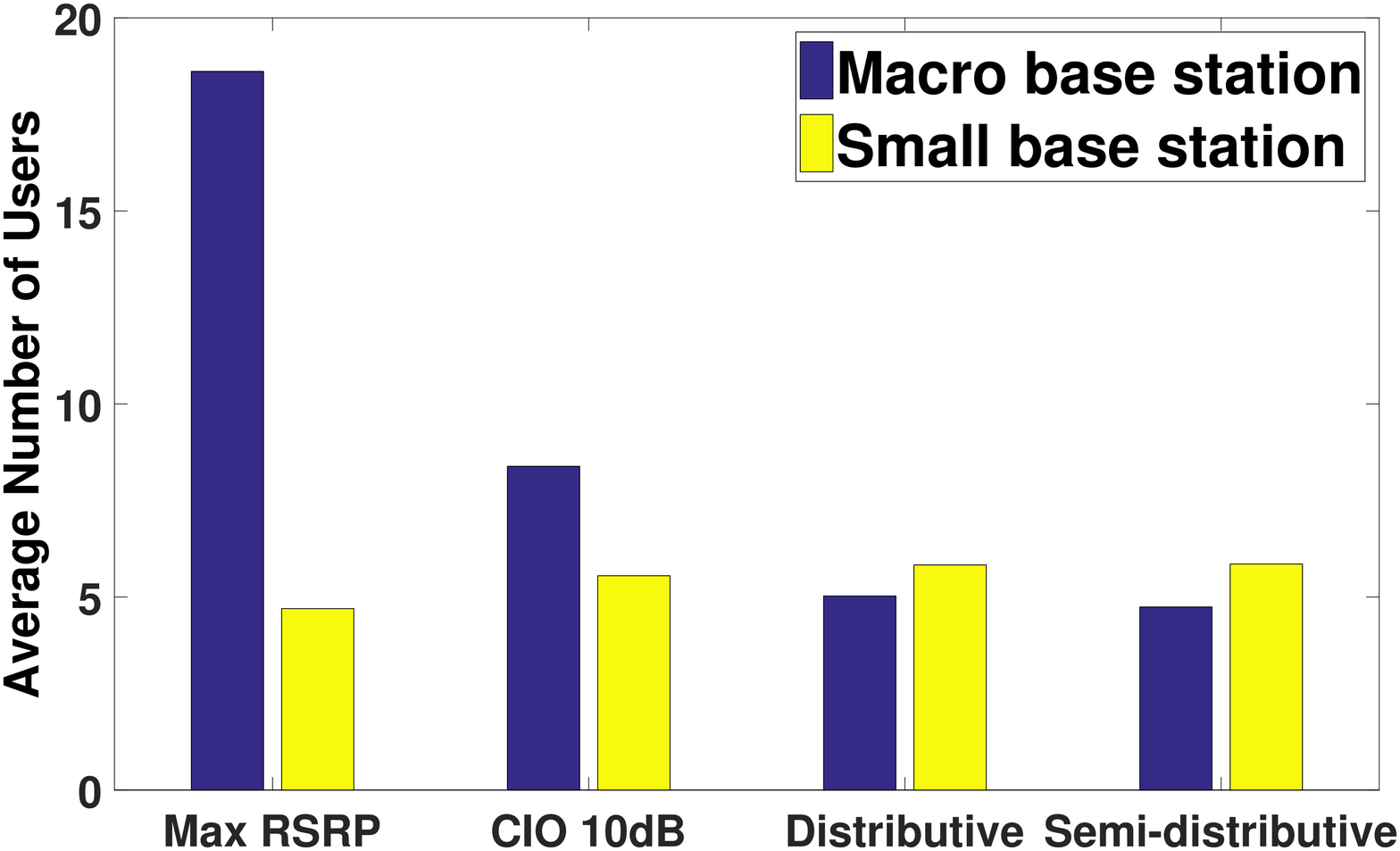}\label{fig:load4SmallBS}}
    \caption{Load distribution per base station among macro and small BS}
    \label{fig:load}
\end{figure}

\textbf{Distribution of Load between Macro and Small BS:} Max RSRP is known to perform poor load distribution between macro and small BS. Macro cells are highly loaded with users as compared to small cells due to transmit power disparity in macro and small cells. The problem of high load on Macro BS is also rectified in our scheme because we give a weight to residual bandwidth. The $AssociationScore$ of equation \ref{eq:associationScore} is dependent on both path loss and available bandwidth for the user which introduces inherent load balancing. Fig. \ref{fig:load1SmallBS} and \ref{fig:load4SmallBS} compare the load distribution on macro BS and small BS with 1 small BS and 4 small BS per sector. Fig. \ref{fig:load4SmallBS} shows that the load is more balanced in our association than both Max SINR and constant CIO based scheme in case of 4 small BS per user. Small BS are slightly more loaded than macro BS in our scheme when there is one small BS per sector. This is because of the fact that more users are concentrated near small BS to model hotspot and only one small BS has more users near it. A close comparison of Fig. \ref{fig:load1SmallBS} and \ref{fig:load4SmallBS} shows that relatively more users are connected to small BS even in Max RSRP and CIO based association when there is just one small BS per sector.

\textbf{Uplink Transmit Power vs Spectrum Efficiency Tradeoff:} There exists an inherent transmit power vs spectrum efficiency tradeoff in our association algorithms. Although user transmit power decreases when we allocate more resources to the user, the spectrum efficiency also decreases. Spectrum efficiency decreases because the same number of bits are being transmitted on a larger chunk of bandwidth. However, both semi-distributive and distributive algorithms allocate a larger chunk of bandwidth to users only when there is residual bandwidth. The availability of residual bandwidth shows low traffic load at a BS. There will be no residual bandwidth at high load and hence there will be no gains in transmit power. Spectrum Efficiency remains the same in high load conditions and spectrum efficiency will only be compromised in low and medium load conditions. 

\textbf{Performance Sensitivity with density of Simulated BS:} The performance sensitivity of our joint user association and resource allocation algorithm with the density of base station is also performed. A comparison of Fig. \ref{fig:power1SmallBS} and Fig. \ref{fig:power4SmallBS} highlights similar gains of our algorithm for both low and high BS density---3 small BS and 12 small BS per cell. The SINR trend also remains the same with varying BS density as highlighted in Fig. \ref{fig:SINR1SmallBS} and Fig. \ref{fig:SINR4SmallBS}. The load distribution among small and macro BS in our scheme improves with an increase in BS density specified in Fig. \ref{fig:load1SmallBS} and Fig. \ref{fig:load4SmallBS}. As explained earlier this is due to the fact that more users are deliberately near small BS for hotspot modeling in our simulations.

\section{Conclusion}
\label{sec:conclusion}
This paper presents a novel resource allocation and user association scheme for user transmit power minimization. This scheme takes into account the residual bandwidth at a base station in addition to path loss. Extra bandwidth is opportunistically allocated to active users for transmit power reduction while meeting user capacity requirements. The performance of proposed method is compared with Max RSRP and CIO based association schemes. Results show a decrease of 22.8 dBm and 11.2 dBm in uplink transmit power for proposed semi-distributive and distributive association algorithms respectively. The reduction in uplink transmit power paves way for including battery constrained IoT devices in 5G and makes a strong case for cellular IoT. 

\section*{Acknowledgment}
This material is based upon work partially funded by the National Science Foundation under Grant Numbers 1619346, 1559483, 1718956 and 1730650. For more information about the projects, please visit www.ai4networks.com

\ifCLASSOPTIONcaptionsoff
  \newpage
\fi

\bibliographystyle{ieeetr}
\bibliography{IoT}

\end{document}